\documentclass[twocolumn,showpacs,showkeys,preprintnumbers,amsmath,amssymb]{revtex4}
\usepackage{graphicx}% Include figure files
\usepackage{dcolumn}% Align table columns on decimal point
\usepackage{bm}% bold math
\usepackage{multirow}

\begin{document}

%%%%%%%%55
\newcolumntype{C}{>{\centering\arraybackslash}p{4em}}
%%%%%%%%

\preprint{}

\title{Universal nature of collective plasmonic excitations in finite 1-D carbon-based nanostructures}

\author{Eric Polizzi}
\email{polizzi@ecs.umass.edu}
\author{Sigfrid Yngvesson}
\affiliation{Department of Electrical and Computer Engineering,
 University of Massachusetts, Amherst}

\date{\today}% It is always \today, today,

\begin{abstract}
Tomonaga-Luttinger (T-L) theory predicts collective plasmon resonances in 1-D nanostructure conductors 
of finite length,
that vary roughly in inverse proportion to the length of the structure.
In-depth quantitative understanding of such resonances which have not been clearly identified in experiments so far,
 would be invaluable for  future generations of nano-photonic and nano-electronic devices 
that employ 1-D conductors.
Here we provide evidence of the plasmon resonances 
in a number of representative 1-D finite carbon-based nanostructures
using first-principle 
computational electronic spectroscopy
studies. 
Our special purpose real-space/real-time all-electron Time-Dependent Density-Functional Theory 
(TDDFT)  simulator can perform excited-states calculations to obtain 
correct frequencies for known optical transitions, and  capture various nanoscopic effects including  collective plasmon excitations.
The presence of 1-D plasmons is universally predicted by
 the various numerical experiments,
which also demonstrate a phenomenon of resonance splitting.
For the metallic carbon nanotubes under study, the plasmons are expected to be related
to the T-L plasmons of infinitely long 1-D structures.
\end{abstract}
%\pacs{02.60.-x,02.70.Hm,02.70.-c,02.10.Ud,31.15.-p,71.15.Dx}% PACS, the Physics and Astronomy

\keywords{1-D conductors, nanoplasmonics, real-time TDDFT, computational electronic spectroscopy, Tomonaga-Luttinger liquid, plasmon velocity}

\maketitle

\section{Introduction}

Nanoplasmonics is a field that has grown rapidly in the last few years \cite{Stockman11,Halas09}. 
Covering a range from terahertz through infrared to visible light frequencies, this field already offers numerous 
applications to electronics and photonics. In the visible and Near IR (NIR) range, nano-antennas \cite{Novotny11} and nanoparticles 
have provided drastically enhanced coupling to electromagnetic waves \cite{Cang11}. 
All of this research work has taken advantage of plasmonic surface modes - either  on nanoparticles, or on metallic sheets -
and has led to some initial practical applications. 
However, there are still significant fundamental limitations that must be overcome for further progress to be possible
 in this field \cite{Plasmonic}.
While there is an extensive literature on 1-D plasmons, 
these have not been emphasized for applications due to the difficulty of preparing devices and performing experiments.
Yet, 1-D conductors are of interest since they
can potentially exhibit lower losses than the 2-D ones.
The chief signature of 1-D plasmons is a high-frequency excitation (i.e. energy resonance) 
that depends inversely on the length of the conductor. 
Examples of nano-structures that  can support 1-D plasmons are graphene nanoribbons (GNRs), 
single wall carbon nanotubes (SWCNTs) and linear carbon chains.  
Due to the very strong coulomb interactions in such 1-D systems,
the low-energy excitations of interacting electrons cannot be adequately described by 
Landau's Fermi liquid theory \cite{Kittel05}, which has to be replaced by Tomonaga-Luttinger (T-L) liquid 
theory \cite{Tomonaga50,Luttinger63,Deshpande10,10-Haldane}.
Specifically, it was shown that metallic carbon nanotubes can be well described with the T-L theory \cite{11-Egger,12-Kane}.
A bosonized representation of the Hamiltonian led to the following conclusions: (i) the
density of single particle states is a power function of the energy and vanishes at the
Fermi energy; (ii) consistent with (i) the conductance of electrons tunneling into the CNT
has a universal power law dependence on voltage and temperature; (iii) the electron
states are separated into charge states and spin states \cite{13-Auslaender}. In the simulations that follow
spin/charge separation effects are neglected; (iv) collective boson states form charge
density waves in the CNT (``T-L plasmons'') which propagate with a velocity $v_P = v_F /g$
where $v_F$ is the Fermi velocity $\simeq 1 \times 10^{6} m/s$, \cite{Saito98} and $g$ is a parameter that depends on the
strength of the Coulomb interaction and on screening by the electrostatic environment
(the latter being a weak effect) \cite{12-Kane} . 
The predictions (i) and (ii) have been well verified by transport 
\cite{16-Bockrath,17-Yao} and optical \cite{18-Ishii} measurements.
The parameter $g$ is $< 1$ for T-L liquids and $=1$ for Fermi
liquids, $v_P$ is thus higher than $v_F$. 
 An average plasmon velocity of the predicted magnitude was recently deduced from absorption measurements on
enriched SWCNTs films \cite{Zhang13}. 
Ultra-fast time-dependent measurements evidenced ballistic electron oscillations in isolated SWCNTs,
 but the propagation velocity was equal to the Fermi velocity, consistent with single particle excitations, 
not plasmons \cite{Zhong08}. 
Detailed experimental confirmation of the predicted 
plasmon resonance frequencies in isolated SWCNTs is thus still an unsolved problem.
Estimates of $g$ for CNTs yield values in the range $0.26-0.33$
\cite{12-Kane} resulting in $v_P$ being in the range $3-4\times 10^6 m/s$. Electromagnetic simulations have yielded
$v_P$ up to $6.2\times 10^6 m/s$ \cite{15-Hanson}.  Recently, T-L liquid behavior with $g=0.53$ was verified in
atomic gold chains by tunneling and optical measurements \cite{19-Blumenstain}.

The focus of this paper is the dynamic behavior of the collective medium in 1-D
conductors and its resonances. A physical picture of the resonances can be gained by
realizing the equivalence of the finite length 1-D conductor to a transmission line \cite{Burke02,21-Bockrath}. The
T-L plasmon waves resonate as they are reflected at the ends of the conductor and the
fundamental resonance frequency will be given by $f = v_p/(2L)$ \cite{22} where $L$ is the length of
the conductor. Single quasi-particles similarly resonate at $f = v_F /(2L)$. In our
simulations we explore both of these types of resonances as well others
that will be discussed in the text.
%The T-L theory predicts two types of resonances: one due to a single particle excitation
%that propagates at the Fermi velocity $v_F$, and another that accounts for 
%collective particles or plasmon excitations \cite{Burke02}. 

%In particular, the Fermi velocity in SWCNTs and related carbon materials is $v_F\simeq 1 \times 10^{6} m/s$ \cite{Saito98},
% while the T-L plasmon velocity $v_{pl}$ predicted for SWCNTs is expected to be much higher 
%(i.e. $\times 3$ to $\times 5v_F$) \cite{Burke02}.

First-principle calculations are becoming critical 
to  provide evidence of plasmon resonances in 1-D carbon-based nanostructures. 
 Such quantitative simulations are known to be challenging, since they 
should be both capable to provide reliable 
information on the many-body excited states (beyond ground state theory), 
and to address large-scale computational needs of finite dimensional systems
 (beyond the solid-state unit-cell).  
The time dependent density functional theory (TDDFT)  \cite{tddft} 
alongside with the simple adiabatic local density approximation (ALDA) for the many-body exchange-correlation term,
 has been very successful for providing accurate absorption spectra of a large number of complex molecular systems \cite{tddftbook}.
The real-time TDDFT approach  introduced by  Yabana and Bertsch \cite{Yabana96,Yabana06}, in particular,
combines both potential for parallel computing scalability, and an intuitive treatment of the 
real-time spectroscopy that can deal with any form of excitations.
Our in-house simulator, named NESSIE, is performing both 
all-electron real-space DFT ground state and real-time TDDFT excited states calculations.
NESSIE benefits from the high-performance capabilities of the
FEAST eigensolver \cite{Polizzi09,feast}, which has also allowed to efficiently redesign various stages of the electronic structure 
numerical modeling
 process \cite{Levin12,Gavin13,Chen10}.
Detailed information on our numerical real-space and real-time modeling framework
is provided %in the method section of this article and  
within the supporting document of this article.

Two types of real-time TDDFT simulations are considered in our experiments. First,  
the spectroscopic information is obtained from linear response calculations 
 after a short and weak polarized impulse is applied along the longitudinal or perpendicular
direction of the 1-D nanostructure. Then, once resonances of interest are identified in the absorption spectrum, 
new time-dependent calculations are performed  in response to a realistic stimulus, such as a laser tuned
 to each resonance frequency (e.g. sinusoidal excitation along the longitudinal direction). 
Such simulations aim at providing more detail on 
the 3-D electron dynamics of the particular resonances with relevant information about their nature.
From atom and benzene-like chain structures to short carbon nanotubes, our numerical experiments discussed below progressively
account for an increase in the physical and computational complexities of the 1-D atomic structure. \\

%\noindent {\bf Carbyne.}
\section{Carbyne}

\begin{figure}
\centering
\includegraphics[width=\linewidth]{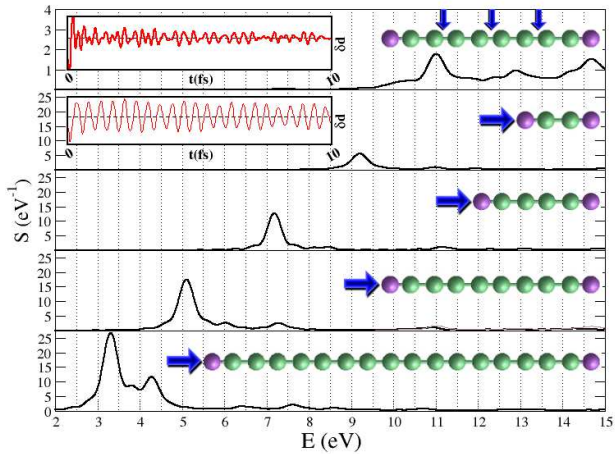}
\caption{\label{fig:chain} Computed absorption spectra of $\rm NC_{2n}N$ with lengths $\rm n=1,2,4,8$ 
(using the single-triple bond alternation structure and the optimized geometries reported in Ref. \cite{Zhang05}).
The plot on the top for $n=4$ is associated with the perpendicular response of a weak impulse applied along 
the same direction. All the other plots consider the response of an excitation applied along the 
longitudinal direction of the chain. The figure insets illustrate 
the variation of the induced dipole obtained from our real-time TDDFT calculations 
and that are used to derive the absorption spectra of the corresponding structures.}
\end{figure}

Carbyne is a chain of carbon atoms that comprises either double or alternating single and triple atomic
bonds. With recent progress in synthesis \cite{Chalifoux10}, first-principle theoretical investigations have become
 increasingly more relevant to study the various attractive physical properties of this material \cite{Artyukhov14}.
Previous studies of carbon chains using the TDDFT approach have been reported in Refs. \cite{Yabana97,Berkus02,Huang14}.
Our numerical simulations of $\rm NC_{2n}N$ shown in Figure \ref{fig:chain},  
are in good agreement with the experimental data found in Ref. \cite{Schermann97}, 
where the strong longitudinal 
resonance  shifts progressively towards the left of the spectrum with longer 
chains (i.e. red shift). 
Their oscillator strength increases with the length of the chain indicating that
 more electrons can participate in the plasmonlike collective excitations.
 Furthermore, the associated
velocity of the resonances is in the typical range of collective plasmons
(as discussed further). References \cite{Yabana97,Huang14}
also identify these excitations as collective plasmons. 
We note that the carbynes are not metallic and thus are not expected to conform with the T-L theory. 
Nevertheless, carbynes show strong 1-D Plasmon excitations.
Moreover, our results show that these longitudinal resonances do not appear with a perpendicular excitation
 (see the $\rm NC_8N$ plot at the top of Figure \ref{fig:chain}). In contrast,  one can identify other small resonances for instance,
at $\rm 11.01 eV$, %$\rm 12.9eV$ 
and at $\rm 14.68 eV$ which do not depend on the length of the chain. 
These particular energy transitions are comparatively close to the ionization potentials of the local atoms 
$\rm C$ and $\rm N$, 
respectively at $\rm 11.26 eV$ and $\rm 14.53 eV$ \cite{Nist}. 

\begin{figure}
\centering
\includegraphics[width=\linewidth]{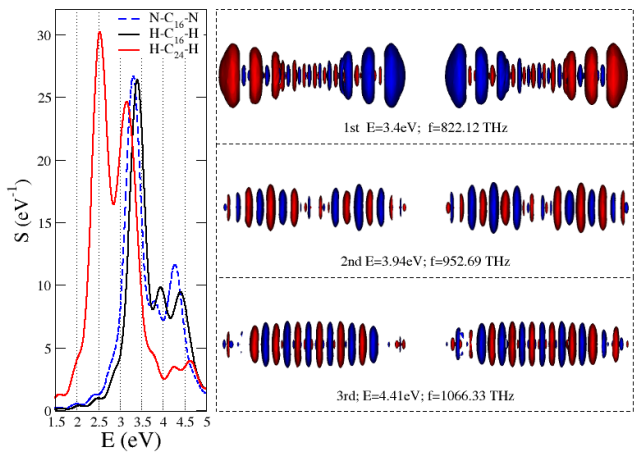}
\caption{\label{fig:chainz} The left plot includes a more detailed representation 
of the main resonances of $\rm NC_{16}N$ observed in Fig. \ref{fig:chain}; it is also plotted alongside
with the absorption spectra of the H-terminated carbon chains $\rm HC_{24}H$ and $\rm HC_{16}H$ (using 
the optimized geometries reported in Ref. \cite{Horny02}). The latter counts
three longitudinal resonances at $\rm 3.4eV$, $\rm 3.94 eV$ and $\rm 4.41 eV$. After new excitations of the structure at each one of these 
specific energy transitions (i.e. sinusoidal excitation at specific frequencies), the figures on the right represent
the 4D isosurface snapshots of the charge oscillation (i.e. the deviation 
of the charge density from the ground state -- positive and negative deviations respectively in red and blue) 
all taken at a different time of the simulation where the dipole goes
 through a maximum and a minimum, respectively.}
\end{figure}

The longitudinal resonances can be observed in more detail in Figure \ref{fig:chainz} (left plot) 
for the $\rm NC_{16}N$, $\rm HC_{16}H$, and $\rm HC_{24}H$ chains. 
We note that the spectra produced  for the $\rm C_{16}$ chain using the H or N termination are very similar. 
Our results are also quantitatively close to the ones obtained in Ref. \cite{Huang14} for the $\rm HC_{2n}H$ chain 
that makes use of the projector augmented-wave (PAW) pseudopotential and performs TDDFT linear response simulations in the frequency domain. 
The lowest energy plasmon resonances found in these results are also in agreement with earlier real-time TDDFT simulations 
performed in Ref. \cite{Berkus02} using a smooth pseudopotential. In comparison with the latter however,
both the PAW and our full-core potential treatments reveal additional plasmonlike resonances at higher energies. 
These oscillations can be observed in the experimental data of the $\rm NC_{n}N$ chains \cite{Schermann97}.
 In our simulations, the splitting between these resonances becomes narrower 
for the $\rm HC_{24}H$ and only two main peaks can be observed instead of three for $\rm HC_{16}H$.
The isosurface snapshot plots in Figure \ref{fig:chainz} represent
three additional real-time TDDFT simulation results 
that can help provide more insights on the time-dependent electron dynamics of the plasmonic resonances in  $\rm HC_{16}H$.
The strongest resonance at the lowest energy transition gives rise to a distinguishable plasmon 
 that can oscillate
back and forth along the longitudinal direction. This collective particle excitation likely takes place 
within the first Van Hove singularity of the 1-D structure where the electrons are strongly coupled.
It has the signature of a  plasmon with a calculated velocity of $v_{pl}=3.18\times 10^6 m/s$ which increases 
to $v_{pl}=3.64 \times 10^6 m/s$ for $\rm HC_{24}H$. 
%Here, we consider $v=2Lf$ where $f$ is the resonance frequency and $L$ is the
%the length of the 1-D structure from the carbon edges.
% length 16 ==> 19.34A with 822.12THz (3.4eV);   length 24 ==> 29.67A with 612.72THz (2.53eV)
The results for the other two snapshot plots confirm the presence of additional 
longitudinal resonance modes 
that were mentioned in Ref. \cite{Huang14},
 with cosine and sine like envelope features for the plasmonic excitations.
These two excitations have a different character in that the charge density alternates from
positive to negative for each atom pair, analogous to optical phonon modes.
Finally, we note that no single particle excitation at the Fermi velocity is observed since the carbon chain family considered here is semiconducting. \\

%\noindent {\bf Acenes and Poly(p-phenylene).}
\section{Acenes and Poly(p-phenylene)}

Acenes and Poly(p-phenylene) (PPP) are
polycyclic aromatic hydrocarbons consisting of linear benzene rings that are respectively fused or attached.
They can also be thought of as the narrowest graphene nanoribbons (GNR) of finite lengths, 
associated with the zig-zag configuration for acenes (2-ZGNR) and the armchair one for PPP (3-AGNR).
Understanding the properties of long acenes in particular, is the subject of active research \cite{Kortya14}.

\begin{figure}
\centering
\includegraphics[width=\linewidth]{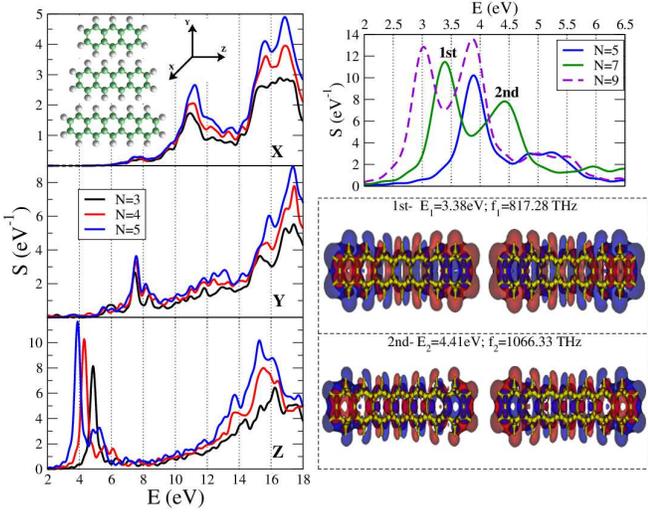}
\caption{\label{fig:zgnr} 
The left plots present the computed absorption spectra of anthracene (N=3), tetracene (N=4), pentacene (N=5) associated
with the responses of weak impulses separately applied along the three main directions 
(regular bond lengths are used for the geometries: $l_{\rm C-C}=1.42\AA$ and $l_{\rm C-H}=1.09\AA$).
The top right plot focuses on the plasmonic response of an excitation along the longitudinal direction
for pentacene (N=5), heptacene (N=7) and nonacene (N=9).  The heptacene, in particular, counts
two main resonances at $\rm 3.38 eV$ and $\rm 4.41 eV$. The 4D isosurface snapshots represent
 the charge oscillations associated with these two energy transitions obtained at different simulation times  
where the dipole response reaches a maximum and a minimum.}
\end{figure}

Results for anthracene, tetracene and pentacene reported in Figure \ref{fig:zgnr} confirm that the strong longitudinal UV resonances 
along the z-direction that appear 
respectively at  $\rm 4.85 eV$,  $\rm 4.30 eV$, $\rm 3.88 eV$,
are in remarkable agreement with the experimental data in Ref. \cite{Freidel51}.
While TDDFT using traditional exchange-correlation functionals such as ALDA fail 
to describe the important
low-lying excited states in acene compounds \cite{Lopata11}, it appears 
 capable to accurately capture the main nanoscopic effects.
These resonances are plasmonlike, since we note a progressive shift towards the left of the spectrum
with longer chains accompanied by an increase in oscillator strength. They are also absent from the results of excitations 
along the $x$ and $y$ directions. For the latter, in turn, one can identify two other resonances
that stay independent of the chain length, at $\rm 11.22 eV$ using an x-polarized impulse that
 equally excites all the carbon atoms in the y-z plane (energy relatively close to the ionization potential of $\rm C$), 
and at $\rm 7.55 eV$ using an y-polarized impulse that 
equally excites all the individual fused benzene rings (energy relatively close to the $\pi\to\pi^*$ transition in benzene).
For the case of pentacene, in addition to a strong plasmon resonance at $\rm 3.88 eV$, we note a couple of
weaker longitudinal resonances at $\rm 4.84 eV$ and $\rm 5.25 eV$ (see top right plot of Figure \ref{fig:zgnr}). 
The latter comes close to the experimental HOMO-LUMO 
gap at $\rm 5.2 eV$ \cite{Sato87} (this gap is known to be severely underestimated by DFT at $\rm \sim1eV$).
The results on heptacene and nonacene in Figure \ref{fig:zgnr}, clearly show that as the length of the structure increases, 
a second strong resonance appears. 
The isosurface snapshots representing the electron dynamics for the two main resonances of
 heptacene 
confirm the presence of two plasmons. 
In contrast to the case of carbon chains, this second resonance does not result from an additional
 longitudinal confinement mode, and the results indicate the presence of another channel 
(i.e. two confinement modes are then present in the transverse $x,y$ plane).
We note that similar double plasmon resonances have also been recently observed using semi-empirical simulations  
applied to GNR structures with variable widths \cite{Cocchi12}.
The calculated plasmon velocities associated with the first and second resonances of heptacene, 
 $v_{{pl}_1} = 2.77 \times 10^6 m/s$ and  $v_{{pl}_2} = 3.61 \times 10^6 m/s$, increase to 
 $v_{{pl}_1} = 3.18 \times 10^6 m/s$ and  $v_{{pl}_2} = 4.08 \times 10^6 m/s$ for nonacene.
%% pentacene length 12.295   e=3.88eV  vp=2.31  --- 4.84eV guess ==> 1170.30 THz, vp2=2.88
%% heptacene length 16.937   e1=3.38 e2=4.41
%% nonacene length  21.776  -e1=3.02 e2=3.87

\begin{figure}
\centering
\includegraphics[width=\linewidth]{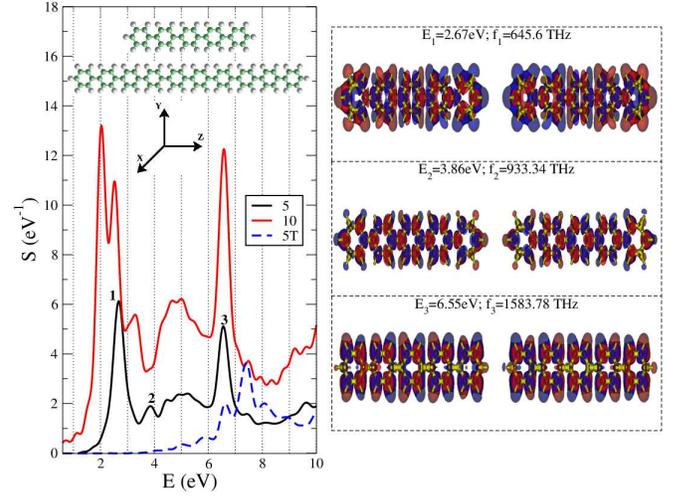}
\caption{\label{fig:ppp} 
The left plot presents the computed absorption spectra of a finite PPP with 5 and 10 unit cells (using the 
geometries reported in Ref. \cite{Vogl95}),
obtained after excitations along the z longitudinal or y perpendicular direction. The result for the latter is noted 5T 
(for the 5 unit cells). The 4D isosurface snapshots on the right represent
 the charge oscillations associated with three main selected longitudinal resonances of the 5 unit cell PPP.}
\end{figure}

The case of finite PPP is shown in Figure \ref{fig:ppp} using lengths of 5 and 10 unit cells.  
Excitations along the perpendicular directions (including the x-direction not shown here) provides results comparable 
to the ones obtained for acenes in Figure \ref{fig:zgnr}. 
However, we note one additional resonance at around $\rm 6.55 eV$ for an excitation along the y perpendicular direction
(curve denoted '5T') which is also present (and amplified) with an excitation along the longitudinal z direction
 for both 5 and 10-PPP. All the other longitudinal resonances at lower energies
are absent from the '5T' curve. The isosurface snapshots of the charge oscillations, 
provide some useful information on the nature of the three main selected longitudinal resonances for 5-PPP.
The strong first peak can be associated with a  plasmon, while the second and third peaks reveal different 
characteristics related to ``band-to-band'' transitions. 
 In particular, the second peak shifts from $\rm 3.86 eV$ to $\rm 3.3 eV$ for the long 10-PPP which is 
in very good agreement with the strong experimental absorption peak (related to the bandgap) found at $\rm 3.4 eV$ for PPP \cite{Tabata86}.
The position of the third peak, in turn, is not sensitive to variation in length. A closer look at the isosurface plots show that
 the charge oscillations at $\rm 6.55eV$ are somehow localized within each attached benzene ring. 
We note that a broader ``band-to-band'' transition region appears at around $\rm 5 eV$ for both  5 and 10-PPP.
Finally, the main plasmon resonance at $\rm 2.67 eV$ i.e. $v_{pl}= 2.44 \times 10^6 m/s$
shifts towards the left of the spectrum of the 10-PPP and, similarly to the case of long acene,
 splits into two peaks. 
The calculated plasmon velocities associated with the two resonances
are  $v_{{pl}_1} = 3.97 \times 10^6 m/s$ and  $v_{{pl}_2} = 4.83 \times 10^6 m/s$. \\

% L=18.8595  5PPP  -- e=2.67eV  ==> vp=2.44m/s
% L=39.8145  10PPP -- e1=2.06eV e2=2.51eV

%\noindent {\bf Single-Wall Carbon Nanotube (SWCNT).}

\section{Single-Wall Carbon Nanotube (SWCNT)}
\begin{figure}
\centering
\includegraphics[width=\linewidth]{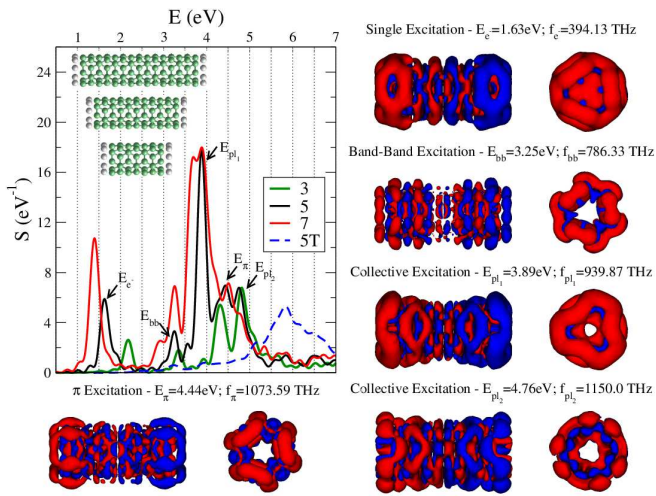}
\caption{\label{fig:cnt} 
The computed absorption spectra of the finite (3,3) armchair SWCNT with three different length 3, 5 and 7 unit cells 
(the C-C bond length is fixed at 1.44\AA~ and one additional carbon ring of type 'A' is added 
to the structures e.g. 'H-AB-AB-AB-A-H' for the 3 unit cells with Hydrogen
 termination). The '5T' curve indicates the result from an excitation in the perpendicular direction, 
all other results consider an excitation along the longitudinal direction of tube.
 The 4D isosurface snapshots on the right and below the spectrum represent the charge oscillations associated with five 
main selected longitudinal resonances of the 5 unit cell tube.
They have all been  taken at a different time of the simulation where the dipole goes
 through a maximum, and the figures include both side and edge views of the nanotube.}
\end{figure}
Figure \ref{fig:cnt} shows our results for simulations of a finite (3,3) armchair SWCNT.  
To stabilize the structure one additional carbon ring, and hydrogen atoms were added at the ends. 
The H atoms are expected to have only minor effects on the 
resonances observed. Armchair SWCNTs are known to exhibit metallic conduction \cite{Saito98}
 and, as
mentioned in the introduction, they can be described by the T-L theory in the infinite
length limit \cite{11-Egger,12-Kane}. We simulated tubes of three different 
lengths, 3, 5, and 7 ``unit cells'', respectively.	
The simulations were performed for an E-field parallel to the z-axis 
of the tubes (the long axis), marked '3', '5' and '7', as well as perpendicular to that axis (for the 5 unit cell case, marked '5T').  
   We first distinguish resonances that do not change with the length (L) of the tube.  The peak at about $\rm 4.44 eV$ 
can be identified as being due to the $\pi$-plasmon ($E_\pi$ in the Figure). 
Note the characteristic distribution of the oscillating charges, following the  bonds, in the snapshot displayed below the spectra. 
A broad band with a peak at $\sim\rm 15 eV$ due to  the $\pi+\sigma$ plasmon can also be observed in the spectra
(observations are included in the supporting information document). 
Since the above plasmons are basically of the bulk (3-D) type, they can also be excited by a field in the direction perpendicular 
to the tube. 
Both of these resonances are well known for 
SWCNTs and have been detected experimentally by Electron Energy Loss (EEL) spectroscopy \cite{Kramberger08} as well as by optical 
spectroscopy (the $\pi$-plasmon) \cite{Kataura99,Murakami05}.  
We note that the $\pi$-plasmon resonance for perpendicular excitation is shifted to a higher energy, $\rm 5.8 eV$, in qualitative 
agreement with the experimental data. 
The $\pi$-plasmon resonance for the 3 unit cell case is not so easy to identify, see further discussion below. 
There is also a resonance for all values of L at about $\rm 3.25 eV$ ($E_{\rm bb}$ in Figure \ref{fig:cnt}). This resonance energy agrees
 with experimental  optical data and simulations \cite{Kataura99,Spataru04} and can be identified as a ``band-to-band'' resonance 
for infinitely long tubes and for higher energy bands. 
The simulations show that this peak does not vary with L, and also that it can not be excited perpendicular to the axis, as expected 
for a band-to band transition.  The charge distribution under sinusoidal excitation shown in Figure \ref{fig:cnt} clearly lacks the collective 
character evident in the plasmon resonances.

One type of resonance that does vary with L occurs in the lowest energy range, from about $1.4$ to $\rm 2.2 eV$ ($E_{e^-}$ in the Figure).  
Calculating a velocity for this resonance we find a value $0.98 \times 10^6 m/s$ and  $1.18 \times 10^6 m/s$ respectively for the 5 and 7
unit cell cases, which is close to the Fermi velocity. We note that this resonance 
can only be excited by a field parallel to the tube axis, as expected if it is due to electrons resonating from one end of the tube 
to the other. This is the type of single particle excitation resonance that was measured in Ref.  \cite{Zhong08}. 
 Further evidence for this interpretation is obtained 
by inspecting the top 4-D isosurface snapshots to the right of the spectra 
('Single Excitation - $E_{e^-} = {\rm 1.63 eV}$'), which show the periodic oscillation of the charge density.

%%%% Lengths %%% Ee
%%%% 3 7.4916     2.18eV  527.13 THz    0.7898
%%%% 5 12.4854    1.63eV  394.13 THz    0.9841
%%%% 7 17.4794    1.4eV   338.52 THz    1.1834

%%%% Lengths %%% Epl
%%%% 3 7.4916     4.32 ``4.81==> pi?''  1044.57 THz 
%%%% 5 12.4854    3.89 4.76 ==>  939.87 THz  1150.0 THz ==>  2.3469 2.87
%%%% 7 17.4794    3.69 3.89 ==>   892.24 THz 939.87 THz ==>  3.119 3.2856 

Finally, we find two strong peaks at energies of $\rm 3.69 - 3.89 eV$ (a split peak for the longest tube) and  $\rm 3.89 eV$  
(marked $E_{{pl}_1}$ in the Figure) that vary with L as expected for T-L plasmon type resonances.  
These resonances can  only be excited with a parallel field as was confirmed 
by a simulation of the 5 unit cell tube with perpendicular excitation (marked '5T').  One of the series of 4-D snapshots 
(‘Collective Oscillation - $E_{{pl}_1} = {\rm 3.89 eV}$’) shows how the charge density waves travel back and forth on the 
tube under sinusoidal excitation at the resonance frequency.
There is a related plasmon resonance at a higher energy ($E_{{pl}_2}$ at $\rm 4.76 eV$) for the 5-unit cell case. 
Thus we find split plasmon resonances for both cases - the 5 unit and 7 unit cell ones - with a much smaller split for the longer tube. 
The calculated velocities for the two T-L plasmons for the 5 unit cell case are 
 $2.35 \times 10^6 m/s$ and  $2.87 \times 10^6 m/s$, and they increase to
$3.12 \times 10^6 m/s$ and  $3.29 \times 10^6 m/s$ for the the 7 unit cell.
The side view snapshots of the charge distribution in Figure \ref{fig:cnt} can now be used to discuss further the different 
characters of the resonances $E_{e^-}$, $E_{{pl}_1}$ and $E_{{pl}_2}$. The charge density for the single (quasi-particle) 
excitation ($E_{e^-}$) is spread across the cross-section. In contrast, the plasmon excitations 
 concentrate the charges around the periphery and at the ends of the tube as expected for a boson-type mode \cite{Tomonaga50,Luttinger63,10-Haldane}. 
The split between the two plasmon frequencies can be interpreted as related to two different modes quantized around the periphery.

Given our interpretation of the peaks from $\rm 3.69 eV$ to $\rm 3.89 eV$ as T-L plasmon resonances, we would expect a similar T-L plasmon 
resonance for the 3 unit cell case at a somewhat higher energy, where we see two peaks. As indicated earlier, 
this brings us to the expected energy of the $\pi$-plasmon, however, and a possible interpretation of the two peaks at 
$\rm 4.32 eV$ and $\rm 4.81 eV$ for the 3 unit cell tube is that the T-L plasmon and the  $\pi$-plasmon are coupled.

\section*{Conclusions}

This paper has simulated molecular absorption spectra employing accurate all-electron real-time TDDFT simulations. 
The range in size has been extended from small molecules such as $\rm C_2H_2$ or benzene to carbon nano-structures 
that are equivalent to 1-D conductors with finite lengths. % from 7 to 24 unit cells.  
 Excellent agreement with experimental spectra is obtained when such data is  available.
We find universal features 
for all structures investigated which include carbon chains, narrow armchair and zigzag graphene nanoribbons (i.e. acenes and PPP), 
as well as short carbon nanotubes.  All structures show collective plasmon oscillations characterized by resonant frequencies 
that vary roughly inversely with the length of the structure.
For the metallic structures (in our case only the SWCNTs) the plasmons are expected to be related
to the T-L plasmons of infinitely long 1-D structures.
\begin{figure}
\centering
\includegraphics[width=\linewidth]{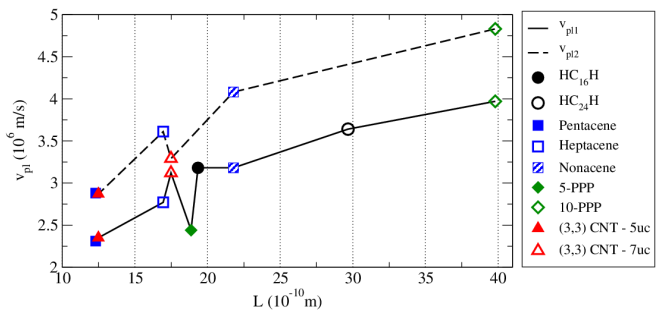}
\caption{\label{fig:table} Summary of plasmon velocities 
computed for all the 1-D carbon nanostructures using different lengths
(numerical values are provided in supporting information document). Two main plasmon modes are represented 
(except for the carbyne and the 5-PPP cases). We note the universal monotonic increase %asymptotic behavior 
of the plasmon velocities for longer structures,
while the less regular behavior in the 15-20\AA~ range (in particular for the 5-PPP and the 7-unit cell 
(3,3) SWCNT) is likely due to coupling effects between ``band-to-band'' transitions and the plasmons.
}
\end{figure}
%%%%%%%%
%In-depth quantitative understanding of such resonances
% would be invaluable for  future generations of nano-photonic and nano-electronic devices that employ 1-D conductors.
%%%%%%%%%%%
Another notable characteristic is that the main plasmon resonance 
is split into two components due to transverse quantum confinement effects
(except for the case of carbyne that presents two to three longitudinal resonance modes).
  The simulations allow vivid 4-D visualizations of the electron charge distributions
 for different types of resonant modes, which then provide more insights on their nature. 
As summarized in Figure \ref{fig:table}, the velocity of plasmon propagation 
shows a monotonic increase with the length of the structure and it 
is expected that the plasmon velocity will asymptotically become independent of length.
Further optimization of our present high performance computing techniques should allow the extension of the
 simulations to  longer structures (up to tens of unit cells for SWCNT), and lead 
 to accurate predicted data that will guide future
photonic and electronic applications of nanostructures over a wide frequency range, from visible to
terahertz.

%\begin{acknowledgement}
%\end{acknowledgement}

\section*{Supplementary information}
%This will usually read something like: ``Experimental procedures and
%characterization data for all new compounds. The class will
%automatically add a sentence pointing to the information on-line:
The supplementary document to this article includes
detailed information about the theoretical models and numerical methods
used in this study, additional information data about the $\pi$ and $\pi+\sigma$ resonances  
in SWCNT, and a table of numerical values for the plasmon velocities.

\vspace{1cm}

%\end{suppinfo}

%\bibliography{achemso-demo}

\begin{thebibliography}{99}
\bibitem{22} The resonance frequency is related to the phase velocity of the resonant wave. For
the basic T-L model the waves are non-dispersive (i.e. the phase velocity is equal to the
group velocity). This is no longer true for the short structures simulated here. 
We derive the plasmon phase velocity 
from the standard formula $v = \omega/k$ where $\omega$ is the angular frequency and $k =2\pi/\lambda$
 is the wavevector. At resonance $L =\lambda/2$. 
 This convention allows us to  show the trend of plasmon velocity
versus length. As the length of the structure increases the dispersion relation 
will asymptotically become linear and the group velocity will become equal to the phase velocity.
\end{thebibliography}

%\section*{References}

\end{document}

%%%%%%%%55
\newcolumntype{C}{>{\centering\arraybackslash}p{3em}}
%%%%%%%%

\noindent {\large {\bf Supporting Information for:}}

~\vspace{3cm}
\begin{center}
\Large{\bf Universal nature of collective plasmonic excitations in finite 1-D carbon-based
nanostructures} \\
Eric Polizzi, Sigfrid Yngvesson \\
Department of Electrical and Computer Engineering, University of Massachusetts, Amherst 
\end{center}

\vspace{3cm}

\tableofcontents

~

\newpage

\section{Methods - Summary}

From physics to algorithms, an account of the multi-step modeling process used by our in-house
real-space and real-time NESSIE simulator is summarized in the following.

\subsection{Ground-state calculations.} 
Our physical model to calculate the ground-state wave functions is at the level of
 Density Functional Theory (DFT) associated with the Kohn-Sham equations. For this study,
the many body exchange-correlation term is defined using a standard local density approximation (LDA) formula \cite{lda}. 
The electronic Hamiltonian accounts for the full core potential (also called all-electron calculations). 
In contrast to pseudopotential methods
in particular, this approach provides higher accuracy, consistency, and higher locality in the
discretization stage with better potentiality for parallelism. 
The number of electrons in our simulations range from a few tens for small molecules
 (e.g. 42 electrons for benzene) to a few hundreds for most of the nanostructures considered here 
(e.g. 552 electrons for the (3,3) SWCNT with 7 unit cells).
The finite element method is used to discretize the Hamiltonian system using tetrahedral elements 
with either quadratic P2 or cubic P3 basis functions (P2 is preferred for our TDDFT 
linear response simulations using weak perturbation -- it is computationally less expensive and  
essentially provides the same results). The real-space simulation domain is taken large enough to minimize any unwanted scattering
from artificial boundaries. More details on our FEM real-space mesh approach are presented in the next Section. %can be found in supplementary information.
In turn, the significant challenges posed by the NESSIE's large-sparse matrix computations resulting from the real-space mesh 
discretization have been addressed recently with the development of our FEAST eigensolver
\cite{Polizzi09,feast}. FEAST has considerably broadened the perspectives for enabling high-performance parallel large-scale calculations.
 In particular, we operate FEAST beyond the current ``black-box'' solver by 
considering a fundamental redesign of the electronic structure numerical modeling, including: 
(i) a reformulation  of the real-space muffin-tin domain-decomposition problem
 for solving the first-principle all-electron problem exactly \cite{Levin12},
(ii) a fundamental non-linear FEAST numerical technique for addressing the self-consistent (SCF) DFT ground-state problem
that represents a robust alternative to traditional SCF methods using 
iterative procedure with density mixing schemes \cite{Gavin13}.

\subsection{Excited-state calculations.} Within the real-time TDDFT framework,
 all the occupied $N_{e}$ ground-state wave functions \mbox{$\{\phi_1 ,  \dots, \phi_{N_e} \}$}
 are propagated (non-linearly) in time by solving a time-dependent Kohn-Sham type equation.
 The absorption spectrum is obtained
 after a short polarized impulse is applied in any given direction $\eta$ (e.g. $x,y$ or $z$) 
of the molecular system i.e. \mbox{$\psi_j(\textbf{r},t=0^+)= \exp(-\imath k\eta) \phi_j(\textbf{r})$} \cite{Yabana06}.
Using the solution of  the electron density at each time step i.e. \mbox{$n({\bf r},t)=2\sum_{j=1}^{N_e}|\psi_j({\bf r},t)|^2$}, 
the induced dipole can then be computed in the time-domain \mbox{$d(t)= \int d{\bf r} n(\textbf{r},t)\eta$},
 and  the imaginary part of  its Fourier 
transform provides  the dipole strength function and the absorption spectrum. Examples on the dipole variations and 
more details on the calculations of the dipole strength function are presented in the next Section. %supplementary information.
The real-time TDDFT technique is often used to address linear responses (e.g. using weak momentum $k$), but the framework
 can also generally address any forms of non-linear responses. In our simulations in particular, 
a longitudinal sinusoidal excitation is often used
 to observe the dynamics of the charge oscillations at a given resonance frequency. 
The real-time propagation is performed 
using either a conventional Crank-Nicolson (CN) scheme or a spectral decomposition technique where
FEAST is used to diagonalize the time-dependent Hamiltonian at each time step \cite{Chen10}. 
This FEAST spectral decomposition technique is well-suited for parallel computing, and it
allows to consider larger time-steps in our simulations $\Delta_t=20as$ (vs $\Delta_t=2as$ using CN).
The total simulation time that has been used varies between 10fs for the carbynes to 15fs for the SWCNTs.

%%%%%%%%%%%%%%%%%%%%%%%%%%%%%%%%%%%%%%%%%%%%%%%%%%%%%%%%%%%%%%%%%%%%%%%%%%%%
%%%%%%%%%%%%%%%%%%%%%%%%%%%%%%%%%%%%%%%%%%%%%%%%%%%%%%%%%%%%%%%%%%%%%%%%%%%%
\section{Methods - Practical considerations  with applications to benzene}

Supplementary information on the modeling procedure used in NESSIE is provided here.
We use the example of the benzene molecule.

\subsection{Real-space discretization}

As specified by an input file, our simulator NESSIE reads the atom coordinates, creates a 3D mesh using tetrahedra, and it builds 
the FEM Hamiltonian and mass (i.e. basis function overlap) matrices using quadratic P2 or cubic P3 precision.

\begin{figure}[htbp]
\centering
\includegraphics[width=\linewidth]{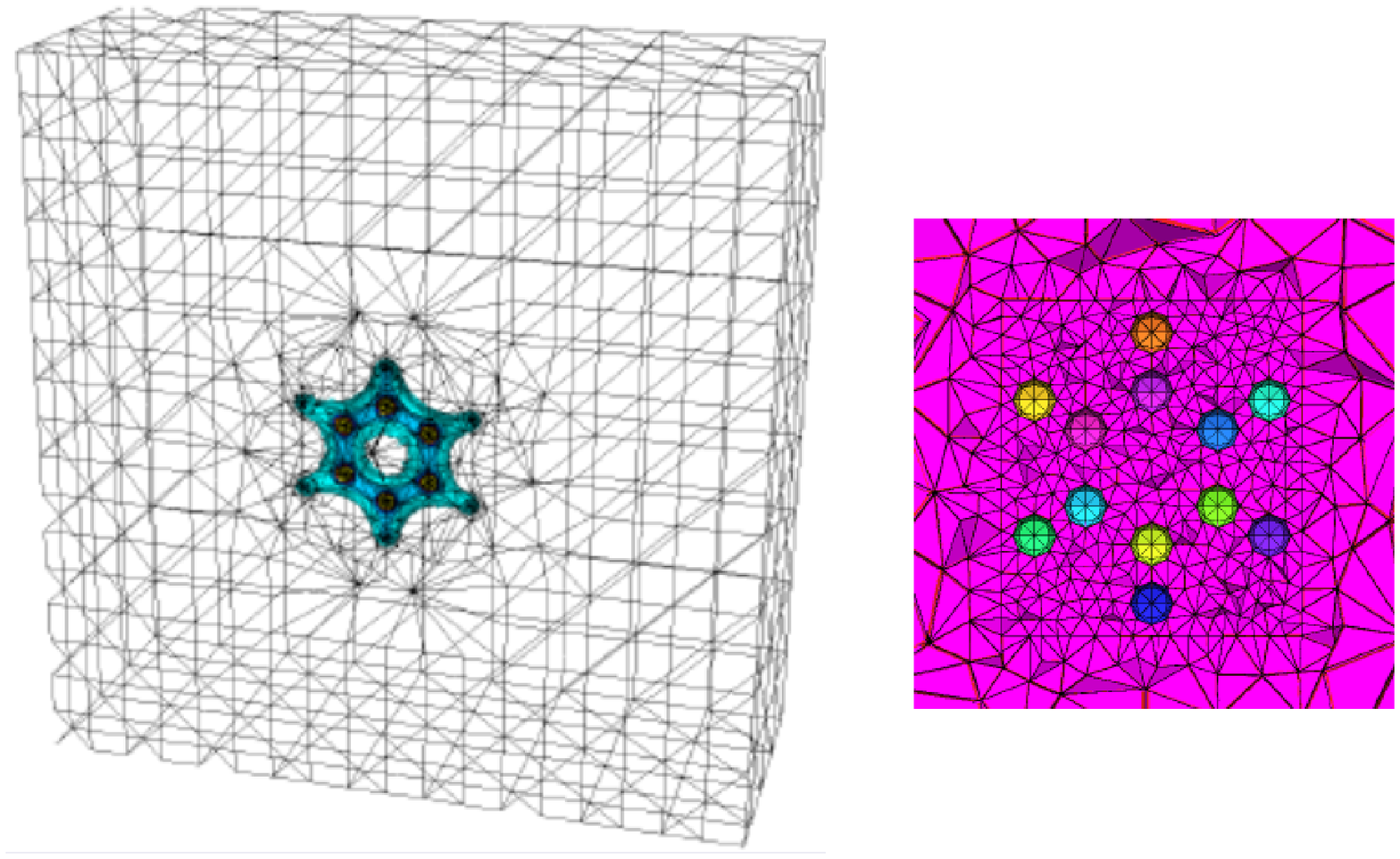}
\caption{\label{fig:muffin} \em Using a muffin-tin domain-decomposition
method, the whole simulation domain $\Omega$ is separated into
multiple atom-centered regions (i.e. muffins on the right) and one
large interstitial region (on the left). The figure in the middle represents a 2D section of local
finite element discretization using a coarse interstitial
mesh (represented only partially here) connecting all
of the atoms of a benzene molecule.}
\end{figure}

The 3D FEM mesh is constructed using a muffin-tin decomposition technique that involves two steps:
(i) a 3D atom-centered mesh which is highly refined around the nucleus to capture the full core potential, 
and (ii) a much coarser 3D interstitial mesh that connects all the atom-centered holes and extends 
to the computational domain boundary. Figure \ref{fig:muffin}
illustrates the essence of our FEM muffin-tin domain decomposition for the
benzene molecule.
The size of the full 3D (reconstructed) sparse system matrix for benzene is $\sim 60K$ using P2, and $\sim 140K$ using P3.
In TDDFT calculations using a weak perturbation, a P2 FEM basis provides results similar to P3.
Table \ref{tab:elec} reports the size of the resulting P2 FEM  Hamiltonian system matrices and the total number of electrons 
for most of the 1-D carbon nanostructures we have considered in our simulations.

\begin{table}[tb]
\begin{small}
\begin{tabular}{|p{6em}|CC|CCC|CC|CC|}
\cline{2-10} %\cline{2-11}
\multicolumn{1}{c|}{}&\multicolumn{2}{c|}{Carbyne} &
\multicolumn{3}{c|}{Acenes} &
\multicolumn{2}{c|}{PPP} &
\multicolumn{2}{c|}{SWCNT}\\
%\hline 
\cline{2-10}
\multicolumn{1}{c|}{} &\multicolumn{1}{c}{$\rm HC_{16}H$} & \multicolumn{1}{c|}{$\rm HC_{24}H$} & \multicolumn{1}{c}{Penta.} & 
 \multicolumn{1}{c}{Hepta.} &  \multicolumn{1}{c|}{Nona.} &  \multicolumn{1}{c}{5PPP} &  \multicolumn{1}{c|}{10PPP}  &  \multicolumn{1}{c}{5-unit} &  \multicolumn{1}{c|}{7-unit} \\ \hline % \cline{2-11}
\# electrons         & 98  & 146  & 146 & 198 & 250 & 202 & 402 &  408 & 552 \\
Size H matrix        & 65,357 & 93,115 & 138,926  & 180,272 & 223,813 &  191,643 & 368,133  & 302,282   & 393,410 \\
%Sparsity \% & 
%\# non-zero elements & 964,673 & 1,375,780 & 2,057,186 & 2,670,122 & 3,315,046 &  2,838,576    & 5,454,258 & 4,497,317 & 5,853,185 \\
\hline 
\end{tabular}
\end{small}
\caption{\label{tab:elec} Hamiltonian system size P2 FEM discretization by NESSIE for various 1-D carbon nanostructures.
The system matrices are very sparse with an average of 15 to 20  non-zero elements per row.} 
\end{table}

The real-space simulation domain $\Omega$ is taken large enough to minimize any unwanted scattering from artificial boundaries.
The wave functions are set equal to zero at the boundaries (i.e. Dirichlet boundary conditions for Kohn-Sham),
and radiative boundary conditions on the electrostatics potential are used for solving the Poisson equation.
The coupled DFT-Kohn-Sham/LDA/Poisson problem is solved self-consistently using the non-linear version of the FEAST solver.
In particular, the lowest $N_e/2$ states are computed ($N_e$ denoting the number of electrons, and the factor 2 accounts for the spin).

\subsection{Real-time calculations and absorption spectra}

Two perturbations of the ground state solutions 
are considered in the simulation (i) a weak short impulse along a given direction (longitudinal or perpendicular 
to the structure), (ii) a weak sinusoidal stimulus at a given resonance frequency.
In both cases, the ground-state occupied wave functions (i.e. electrons) need to evolve in time. 
At each new time step $t+\Delta_t$, the Poisson equation is solved, and the time-dependent ALDA Hamiltonian is updated.
We are also making use of a simple corrector-predictor schemes to evaluate the Hamiltonian at $t+\Delta_t/2$.
For the real-time propagation, two strategies have been used: (i) a Crank-Nicolson (CN) scheme 
that requires solving a linear system at each time step (number of right-hand-side is equal to $N_e/2$);
(ii) a spectral decomposition scheme that requires the partial diagonalization of the large Hamiltonian
at each time step. The latter direct diagonalization technique can  become  
 a viable alternative to CN (i.e. potentially capable of both
higher-scalability and better accuracy) only if used within a parallel computing environment and by taking advantage
of various intrinsic properties of the FEAST solver (e.g. the eigenvector subspace computed in the current time-step can be
 used as a good initial guess for computing the next one).

Thereafter, at each time step we compute the 3d electron density (sum of amplitude square of the occupied wave
functions that are propagated). The variation of the induced dipole is
then computed relative to the dipole of the ground state and the center of mass $r_0\equiv(x_0,y_0,z_0)$ of the nanostructure.
For example, the response along the longitudinal $z$ direction is given by:

$$\delta d_z(t)= \int_\Omega d{\bf r} (n(\textbf{r},t)-n(\textbf{r},0)) (z-z_0).$$

\begin{figure}[htbp]
\centering
\includegraphics[width=\linewidth]{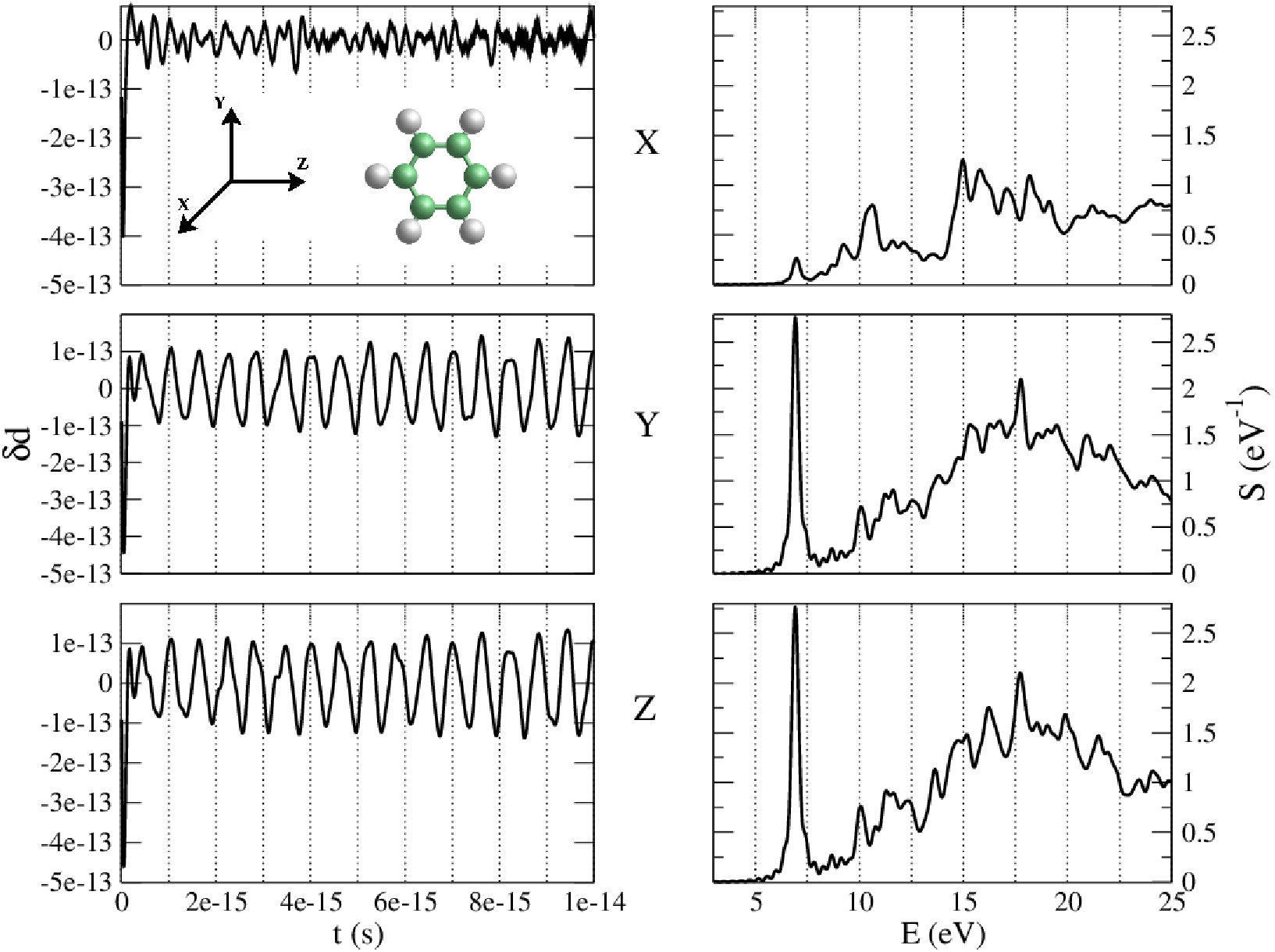}
\caption{\label{fig:c6h6p} \em Dipole variations and corresponding absorption spectrum after
excitation along the three main direction of the benzene molecule. The CN scheme is used for real-time propagation with
time-step $\rm \Delta_t=5as$ and a total time of $\rm 10fs$ (2000 time steps).}
\end{figure}

\begin{figure}[htbp]
\centering
\includegraphics[width=\linewidth]{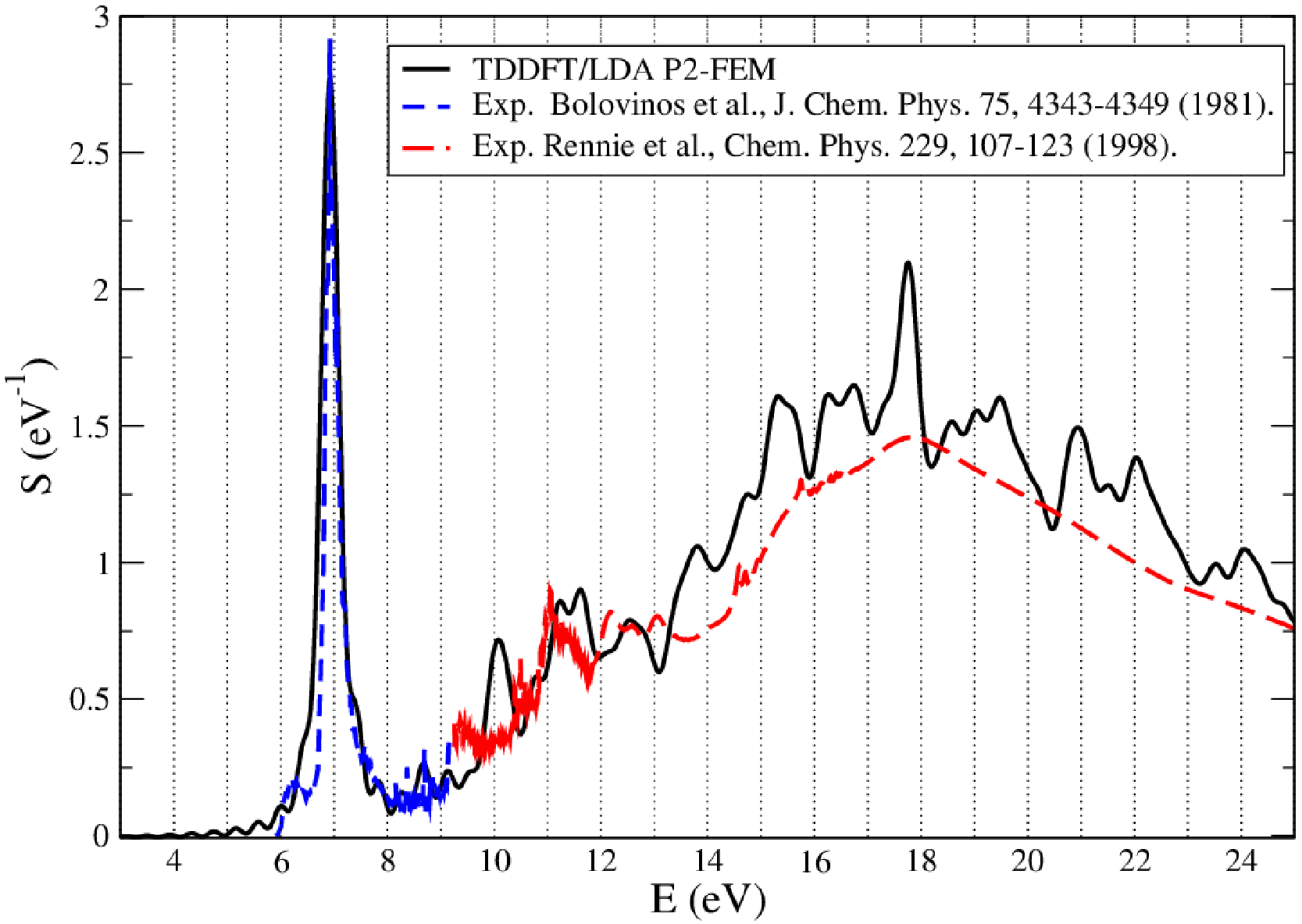}
\caption{\label{fig:c6h6exp} \em Calculated absorption spectrum of the benzene molecule using NESSIE and comparison with experimental data.}
\end{figure}

For the case of a short polarized impulse initial perturbation (using a weak momentum $k$), 
the dipole strength function $S$ is obtained by taking the imaginary part of the Fourier transformation of the dipole moment, i.e.

$$S(E)=\frac{2mE}{\pi k\hbar^3}\Im\left\{\int_0^Tdt \ \delta d(t)e^{-\imath Et/\hbar}w(t)\right\},$$
where $w(t)$ represents a damping factor that introduces artificial dissipation for suppressing the noise 
in the Fourier transform (we choose $w(t)=e^{-\gamma t/T}$). Typically in our simulations,
 we set $\gamma=3$, and $k=0.01/L$ where $L$ is the dimension of the computational domain along the polarization direction.
For the example of the benzene molecule,  Figure  \ref{fig:c6h6p} presents %considers a
the results obtained for the dipole variations and their associated absorption spectrum, after
an excitation is applied along the three main directions of the molecule. 
The mean of the three solutions is plotted in Figure \ref{fig:c6h6exp} and it can be quantitatively compared with the 
experimental data.

\newpage

%%%%%%%%%%%%%%%%%%%%%%%%%%%%%%%%%%%%%%%%%%%%%%%%%%%%%%%%%%%%%%%%%%%%%%%%%%%%
%%%%%%%%%%%%%%%%%%%%%%%%%%%%%%%%%%%%%%%%%%%%%%%%%%%%%%%%%%%%%%%%%%%%%%%%%%%%
\section{Supplementary Figure for SWCNT}

Figure 5 shows the simulation results for a finite (3,3) armchair SWCNTs including 
the $\pi+\sigma$ bulk-type plasmon peak - broad band at $\sim$ 15eV.
We note that both $\pi$ and $\pi+\sigma$ resonances for a perpendicular excitation are shifted to higher energy values 
in qualitative agreement with the experimental data (e.g. Kramberger C. et al., Phys. Rev. Lett., 100,
119803, 2008).

\begin{figure}[htbp]
\centering
\includegraphics[width=\linewidth]{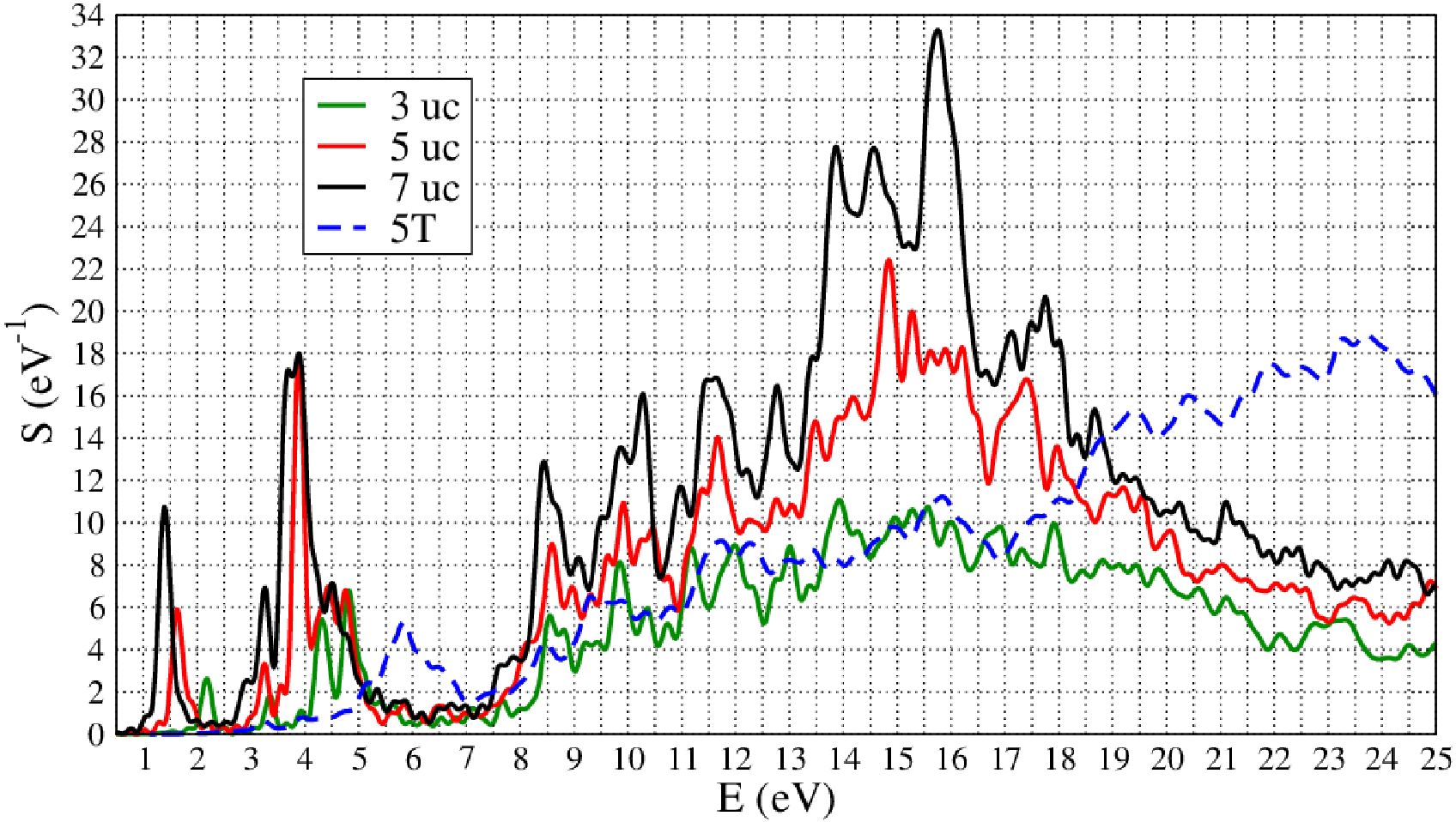}
\caption{\label{fig:c33sigma} \em Calculated absorption spectra of the finite (3,3)
armchair SWCNT with three different length 3, 5 and 7 unit
cells (the C-C bond length is fixed at 1.44\AA~ and one 
additional carbon ring of type ’A’ is added to the structures e.g. ’H-AB-AB-AB-A-H’ for the 3 unit cells with Hydrogen 
termination). The ’5T’ curve indicates the result from an 
excitation in the perpendicular direction, all other results consider
an excitation along the longitudinal direction of tube.}
\end{figure}

%%%%%%%%%%%%%%%%%%%%%%%%%%%%%%%%%%%%%%%%%%%%%%%%%%%%%%%%%%%%%%%%%%%%%%%%%%%%
%%%%%%%%%%%%%%%%%%%%%%%%%%%%%%%%%%%%%%%%%%%%%%%%%%%%%%%%%%%%%%%%%%%%%%%%%%%%
\section{Supplementary Table for Plasmon velocities}

\begin{table*}[htb]
\begin{small}
\begin{tabular}{|p{6em}|CC|CCC|CC|CC|}
\cline{2-10} %\cline{2-11}
\multicolumn{1}{c|}{}&\multicolumn{2}{c|}{Carbyne} &
\multicolumn{3}{c|}{Acenes} &
\multicolumn{2}{c|}{PPP} &
\multicolumn{2}{c|}{SWCNT}\\
%\hline 
\cline{2-10}
\multicolumn{1}{c|}{} &\multicolumn{1}{c}{$\rm HC_{16}H$} & \multicolumn{1}{c|}{$\rm HC_{24}H$} & \multicolumn{1}{c}{Penta.} & 
 \multicolumn{1}{c}{Hepta.} &  \multicolumn{1}{c|}{Nona.} &  \multicolumn{1}{c}{5PPP} &  \multicolumn{1}{c|}{10PPP}  &  \multicolumn{1}{c}{5-unit} &  \multicolumn{1}{c|}{7-unit} \\ \hline % \cline{2-11}
L (\AA) & 19.34  & 29.67  & 12.29 & 16.94 & 21.78 & 18.86 & 39.81 &  12.48 & 17.48 \\
$E_{{pl}_1}$ (eV) & 3.40 & 2.53 & 3.88 & 3.38 & 3.02 & 2.67 & 2.06 &  3.89 & 3.69 \\
$v_{{pl}_1}$ ($10^6m/s$) & 3.18 & 3.64 & 2.31 & 2.77 & 3.18 & 2.44 & 3.97 &  2.35 & 3.12 \\
$E_{{pl}_2}$ (eV) &  &  & 4.84 & 4.41 & 3.87 &     & 2.51 &  4.76 & 3.89 \\
$v_{{pl}_2}$ ($10^6m/s$) &  &  & 2.88 & 3.61 & 4.08 &   & 4.83 &  2.87 & 3.29 \\
\hline 
\end{tabular}
\end{small}
\caption{\label{tab:vpp}
Numerical values of T-L plasmon velocities computed for all the 1-D carbon nanostructures using different lengths.
Two main plasmon modes are reported (except for the carbyne and the 5PPP cases). }
\end{table*}